\documentclass[twocolumn,english,prl,showpacs]{revtex4}
\usepackage[T1]{fontenc}
\usepackage[latin9]{inputenc}
\usepackage{amsmath}
\usepackage{amssymb}
\usepackage{graphicx}
\usepackage{esint}

\makeatletter
\@ifundefined{textcolor}{}
{%
 \definecolor{BLACK}{gray}{0}
 \definecolor{WHITE}{gray}{1}
 \definecolor{RED}{rgb}{1,0,0}
 \definecolor{GREEN}{rgb}{0,1,0}
 \definecolor{BLUE}{rgb}{0,0,1}
 \definecolor{CYAN}{cmyk}{1,0,0,0}
 \definecolor{MAGENTA}{cmyk}{0,1,0,0}
 \definecolor{YELLOW}{cmyk}{0,0,1,0}
 }

\makeatother

\usepackage{babel}
\begin{document}

\title{Phase velocity and particle injection \\
in a self-modulated proton-driven plasma wakefield accelerator}

\author{A. Pukhov$^{1}$, N. Kumar$^{1}$, T. Tückmantel$^{1}$, A. Upadhyay$^{1}$,
K. Lotov$^{2}$, P. Muggli$^{3}$, V.~Khudik$^{4}$, C.~Siemon$^{4}$,
and G.~Shvets$^{4}$}

\affiliation{$^{1}$Institut für Theoretische Physik I, Universität Düsseldorf,
40225 Germany}

\affiliation{$^{2}$Budker Institute of Nuclear Physics and Novosibirsk State
University, 630090 Novosibirsk, Russia}

\affiliation{$^{3}$Max-Plank-Institut für Physik, 80805 München , Germany}

\affiliation{$^{4}$The University of Texas at Austin, Department of Physics and
Institute for Fusion Studies, Austin TX 78712, U.S.A.}
\begin{abstract}
It is demonstrated that the performance of the self-modulated proton
driver plasma wakefield accelerator (SM-PDPWA) is strongly affected
by the reduced phase velocity of the plasma wave. Using analytical
theory and particle-in-cell simulations, we show that the reduction
is largest during the linear stage of self-modulation. As the instability
nonlinearly saturates, the phase velocity approaches that of the driver.
The deleterious effects of the wake's dynamics on the maximum energy
gain of accelerated electrons can be avoided using side-injections
of electrons, or by controlling the wake's phase velocity by smooth
plasma density gradients. 
\end{abstract}

\pacs{52.40.Mj, 52.65.-y, 52.59.-f}

\maketitle
A plasma is a promising medium for high gradient acceleration of charged
particles. It can sustain fields orders of magnitude higher than the
breakdown fields of conventional accelerators \cite{tajima dawson}.
One can excite strong plasma waves either with lasers or with charged
particle beams \cite{joshi malka,joshi scientific american}. One
of the very attractive approaches is to use already existing TeV proton
beams as a driver to generate plasma wake fields. Due to the limitation
set by the transformer ratio, the energy gain of the witness beam
cannot be much larger than the driver energy \cite{TR}. Employing
a TeV proton driver allows in principle to accelerate an electron
bunch to TeV energies in one single stage thus alleviating the technical
burden of multistaging.

It has recently been shown using detailed simulations \cite{pwa,lotov}
that a high gradient plasma wake fields can be generated with an ultra-short
bunch of protons. In that scenario, the proton bunch was shorter than
the plasma wavelength. Unfortunately, such ultra-short proton bunches
are not presently available. The length of existing TeV-class proton
bunches is of order $L\sigma_{z}\sim10$cm. The characteristic plasma
field, the so called wave breaking field is $E_{\textup{WB}}=mc\omega_{p}/e\approx96\sqrt{n_{e}({\rm cm}^{-3})}$V/m,
where $\omega_{p}=\sqrt{4\pi n_{e}e^{2}/m}$ is the plasma frequency
defined by the electron density $n_{e}$. Accelerating gradients of
a GV/m-scale require a plasma density of at least $n_{e}=10^{14}$cm$^{-3}$
corresponding to the plasma wavelength $\lambda_{p}\equiv2\pi/k_{p}\equiv2\pi c/\omega_{p}\sim3$mm.
Thus, the existing proton bunches correspond to $\sigma_{z}=(10-100)\lambda_{p}$
and cannot efficiently generate wake fields in such plasma. The situation
with the proton bunches is very much the same as it was with laser
pulses in the 1980's. The availability of long laser pulses necessitated
the invention of a self-modulated laser wakefield accelerator (SM-LWFA)\cite{esarey sm}.
Subsequent progress in ultra-short pulse laser technology removed
the need for self-modulation and led to successful mono-energetic
electron acceleration in the bubble regime \cite{bubble} that reached
GeV energies.

A long proton bunch propagating in an overdense plasma is also subject
to self-modulation at the background plasma wavelength \cite{kumar sm}.
The effect of self-modulation opens a possibility to use existing
proton bunches for large amplitude wake field excitation. An experimental
program is currently under consideration at CERN. The Super Proton
Synchrotron (SPS) bunch with $450$ GeV protons is proposed as the
driver for the initial stage of the experimental campaign. The wake
field will be used for accelerating externally injected electrons.
The injected particles must be trapped in the wake field. The trapping
condition depends on the wake field amplitude and phase velocity \cite{esarey review}.
Because it is expected that the SPS bunch will generate a weakly-nonlinear
plasma wave with the same phase velocity $\upsilon_{{\rm ph}}$ as
the speed of the driver, it is natural to assume that the gamma-factor
of the injected electrons $\gamma_{{\rm el}}$ must be comparable
to that of the proton driver $\gamma_{{\rm p}}$ for them to be trapped.
As demonstrated below, that is not the case because the spatio-temporal
nature of the self-focusing instability of the proton bunch considerably
reduces $\upsilon_{{\rm ph}}$.

Although it has been realized for some time \cite{andreev vph} that
the phase velocity of the plasma wake produced by the self-modulation
instability of a laser pulse is slower than the pulse's group velocity,
this was not an important issue because the laser group velocity was
usually modest. For the self-modulated proton-driven plasma wakefield
accelerator (SM-PDPWA), the wake slowdown is of critical importance.
Here we show that the phase velocity of the unstable wave is defined
not so much by the driver velocity, but mainly by the instability
growth rate. The wake field is greatly slowed down at the linear instability
stage when the growth rate is at its maximum. At the nonlinear saturation
stage, the wake reaches the driver phase velocity. We also propose
a method to manipulate the wake phase velocity by smooth longitudinal
density gradients.

To describe the wake slowdown analytically, we adopt the formalism
developed~ within the framework of the envelope description of the
driver \cite{kumar sm}. We assume an axisymmetric bunch driver and
utilize the co-moving and propagation distance variables $\xi=\beta_{0}ct-z$
and $\tau=z/c$, respectively, where $\beta_{0}=\upsilon_{z}/c$ ($\upsilon_{z}$
is the velocity of the bunch) and $z$ is the bunch propagation direction
is $z.$ Further, the driver bunch is assumed stiff enough so that
its evolution time is slow $\partial_{\tau}\ll c\partial_{\xi}$.
The bunch is assumed to be long: $\omega_{p}\sigma_{z}/c\gg1$. The
assumed bunch density profile is $\rho(r,\xi)=\rho_{0}\,\psi(r)f(\xi)$,
where $\rho_{0}=n_{b}e$ is the charge density of the proton bunch.
For simplicity, the step-like radial profile $\psi(r)=\Theta(r_{b}-r)$
is assumed, where $r_{b}(\xi,\tau)$ is the evolving radius of the
bunch's envelope, and $\Theta(r_{b}-r)$ is the Heaviside step-function.
The betatron frequency and wavenumber of the self-focused bunch are
defined as $\omega_{\beta0}\equiv ck_{\beta0}=\sqrt{4\pi n_{b}e^{2}/2\gamma_{b}m_{p}}$,
where $m_{p}$ is the proton mass. In the limit of a thin bunch $(k_{p}r_{b}\ll1)$
and linear plasma response, the equation of motion for the bunch's
radius (in normalized coordinates $\tilde{r}_{b}=r_{b}/r_{b0},\,\tilde{\tau}=\omega_{\beta0}\tau,\,\tilde{\xi}=k_{p}\xi$)
is given by \cite{kumar sm}

\begin{equation}
\frac{\partial^{2}\tilde{r}_{b}(\tilde{\xi})}{\partial\tilde{\tau}^{2}}-\frac{\tilde{\epsilon}_{n}^{2}}{\tilde{r}_{b}^{3}(\tilde{\xi})}=-\int_{0}^{\tilde{\xi}}\frac{I(\tilde{\xi}')}{\tilde{r}_{b}(\tilde{\xi}')}\,\text{sin}(\tilde{\xi}-\tilde{\xi}')d\tilde{\xi}',\label{eq:1}
\end{equation}
 where $\tilde{\epsilon}_{n}=\sigma_{\theta}/(k_{\beta0}r_{b0})$
with $\sigma_{\theta}$, $r_{b0}$ being the bunch angular divergence,
initial radius, and longitudinal current profile, respectively, and
$I(\xi)=f(\xi)\tilde{r}_{b}^{2}(\tilde{\xi},\tilde{\tau})$. Perturbing
Eq.\eqref{eq:1} about the initial radius $\tilde{r}_{b}=1+\delta r_{b}$,
yields the linearized equation \cite{kumar sm}:

\noindent 
\begin{equation}
\left(\frac{\partial^{2}}{\partial\tilde{\xi}^{2}}+1\right)\left(\frac{\partial^{2}\delta r_{b}}{\partial\tilde{\tau}^{2}}+3\tilde{\epsilon}_{n}^{2}\delta r_{b}\right)=\delta r_{b}.\label{pert5}
\end{equation}

Following the approach of Bers \cite{bers}, we find an asymptotic
solution of this equation for sufficiently late times, $\tilde{\tau}>L_{e}$,
where $L_{e}\sim1/\Gamma$ is the $e$-folding length, and $\Gamma$
is the growth rate of the instability. Substituting of $\delta r_{b}=\delta\hat{r}_{b}\,\text{exp}(-i\delta\omega\tilde{\tau}+ik\tilde{\xi})$
into Eq.~(\ref{pert5}) yields the dispersion equation $\delta\omega(k)$.
The peak growth rate is calculated~\cite{bers} by introducing $\delta\omega^{'}=\delta\omega-\upsilon k$,
where $\upsilon=\tilde{\xi}/\tilde{\tau}$, and requiring that $D(\delta\omega^{'},k)=0$
and $\partial D(\delta\omega^{'},k)/\partial k=0$: 
\begin{eqnarray}
(1-k^{2})(-\delta\omega^{2}+3\tilde{\epsilon}_{n}^{2}) & = & 1,\label{eq:disD}\\
-k(-\delta\omega^{2}+3\tilde{\epsilon}_{n}^{2})-\delta\omega\upsilon(1-k^{2}) & = & 0.\label{eq:dispdD}
\end{eqnarray}
Equations (\ref{eq:disD}) and (\ref{eq:dispdD}) lead to the standard
dispersion relation typical of the beam-breakup instability which
is known~\cite{whittum_prl91} to always possess a growing mode with
$\Im(\delta\omega)>0$. To simplify the algebra, we assume that $3\tilde{\epsilon}_{n}^{2}$
is small and consider the initial stage of the instability corresponding
to $k=1+\delta k$, where $\delta k\ll1$. In this limit $\delta\omega\gg\tilde{\epsilon}_{n}$,
and Eqs.~\eqref{eq:disD},\eqref{eq:dispdD} reduce to the dispersion
relations $\delta\omega^{3}=-\upsilon,$ and $\delta\omega=-2v\delta k$.
The complex roots are given by 
\begin{eqnarray}
\delta\omega_{\pm}=\frac{1\pm i\sqrt{3}}{2}\left(\frac{\tilde{\xi}}{\tilde{\tau}}\right)^{1/3},\;\delta k=-\frac{1\pm i\sqrt{3}}{4}\left(\frac{\tilde{\tau}^{2}}{\tilde{\xi}^{2}}\right)^{1/3}.\label{eq:unstable modes}
\end{eqnarray}
 In physical units, the condition for $\delta\omega\gg\tilde{\epsilon}_{n}$
can be expressed as 
\begin{equation}
t\ll t_{{\rm short}}\equiv{\gamma_{b}^{2}}\frac{\xi}{c}\frac{n_{b}m}{n_{e}m_{p}}\left(\frac{k_{p}r_{b0}^{2}}{\varepsilon_{n}}\right)^{3},\label{eq:unstable time}
\end{equation}
where $\varepsilon_{n}=\gamma_{b}\sigma_{\theta}r_{b}$ is the normalized
beam emittance. The number of e-foldings is given by $N_{e}=\Im(\delta\omega)\tilde{\tau}-\Im(\delta k)\tilde{\xi}=3\Im(\delta\omega)\tilde{\tau}/2$,
and therefore the growth rate $\Gamma=3\Im(\delta\omega)/2=3\sqrt{3}\left(\tilde{\xi}I(\tilde{\xi})/\tilde{\tau}\right)^{1/3}/4$.
In dimensional variables, the instability growth rate is expressed
as 
\begin{align}
\Gamma= & \frac{3\sqrt{3}}{4}\omega_{p}\left(\frac{n_{b}m}{2n_{e}m_{p}\gamma_{b}}\frac{\xi}{ct}\right)^{1/3},\label{eq:earlygrow}
\end{align}
and the maximum number of e-foldings achieved at $t=t_{{\rm short}}$
is $N_{e}\approx(3\sqrt{3}/4)\tilde{\xi}^{1/3}\tilde{\tau}^{2/3}=(3\sqrt{3}/4)k_{p}\sigma_{z}\tilde{\epsilon}_{n}^{-2/3}$.
Note that for the typical parameters of the SPS bunch and plasma density
$n_{e}=10^{14}\ldots10^{15}$cm$^{-3}$, $N_{e}\approx100$. Therefore,
it is most likely that the self-modulation instability will enter
the nonlinear regime prior to $t=t_{{\rm short}}$, and the above
assumptions will remain valid throughout the linear stage of the instability.

The crucial observation is that $\delta\omega$ and $\delta k$ have
not only imaginary parts responsible for instability growth, but also
real parts. It is these real parts that change the wake phase velocity.
The wake phase is $\psi_{e}=(1+\Re(\delta k))\tilde{\xi}-\Re(\delta\omega)\tilde{\tau}$,
and the phase velocity $\upsilon_{\mathrm{ph}}=\upsilon_{b}+(\partial\psi/\partial\tilde{\tau})/(\partial\psi/\partial\tilde{\xi})\omega_{\beta0}/\omega_{p}$.
Substituting \eqref{eq:unstable modes} for $\delta\omega$ and $\delta k$
and neglecting small terms on the order of $\upsilon^{-2/3}$, we
obtain the phase velocity of the growing mode

\begin{equation}
\upsilon_{\mathrm{ph}}=\upsilon_{b}\left[1-\frac{1}{2}\left(\frac{\xi}{ct}\right)^{1/3}\left(\frac{n_{b}m}{2n_{e}m_{p}\gamma_{b}}\right)^{1/3}\right].\label{eq:Vph}
\end{equation}
 The wake phase velocity \eqref{eq:Vph} can be significantly lower
than the speed of the bunch due to the instability dispersion. The
relativistic $\gamma-$factor of the wake phase velocity $\gamma_{ph}=\left(1-\upsilon_{ph}^{2}/c^{2}\right)^{-1/2}$
can be an order of magnitude lower than that of the driving bunch.
This effect will prohibit electron acceleration to high energies at
the growing instability stage. Yet, one can easily see from the formulas
that the phase velocity decrease is closely connected to the instability
growth rate. Thus, one may expect that when the instability saturates,
the phase velocity of the wake becomes close to that of the bunch.
This effect might help to inject low energy electrons into the wake
of a highly relativistic proton bunch at a later stage of the instability,
just before the nonlinear saturation of the instability.

We should mention here that the dispersion relation allows also for
a purely oscillating mode with $\delta\omega_{0}=-\left(\tilde{\xi}/\tilde{\tau}\right)^{1/3},\;\delta k=\frac{1}{2}\left(\tilde{\tau}/\tilde{\xi}\right)^{2/3}.$
This mode has the superluminous phase velocity 
\begin{equation}
\upsilon_{0\mathrm{ph}}=\upsilon_{b}\left[1+\left(\frac{\xi}{ct}\right)^{1/3}\left(\frac{n_{b}m}{2n_{e}m_{p}\gamma_{b}}\right)^{1/3}\right].\label{eq:Vph-fast}
\end{equation}
Stable propagation of a modulated beam has been observed recently
in simulations \cite{Lotov2011}.

The envisioned experimental program at CERN will use the SPS bunch.
It normally delivers $1.15\cdot10^{11}$ protons at 450 GeV/c with
the normalized emittance $\epsilon_{n}=3.5\mu$m and the length $\sigma_{z}=12$cm.
We use these bunch parameters in our 3D PIC simulations with the newly
developed hybrid code H-VLPL3D \cite{tobi2010}. This new code simulates
the background plasma hydrodynamically while high energy bunches are
treated with a full kinetic algorithm. The hydrodynamic part of the
code introduces much less numerical dispersion into the plasma waves
than a PIC code with the same resolution. The bunch focused to $\sigma_{r}=0.19\,$mm
is sent through plasma with the free electron density $n_{e}=7.76\cdot10^{14}\,$cm$^{-3}$.
The maximum bunch density on axis is $n_{b}=1.5\cdot10^{12}\,$cm$^{-3}$.
To avoid beam hosing and to seed the self-modulation, we assumed the
bunch is hard-cut in the middle \cite{kumar sm}.

\begin{figure}
\includegraphics[width=0.9\columnwidth]{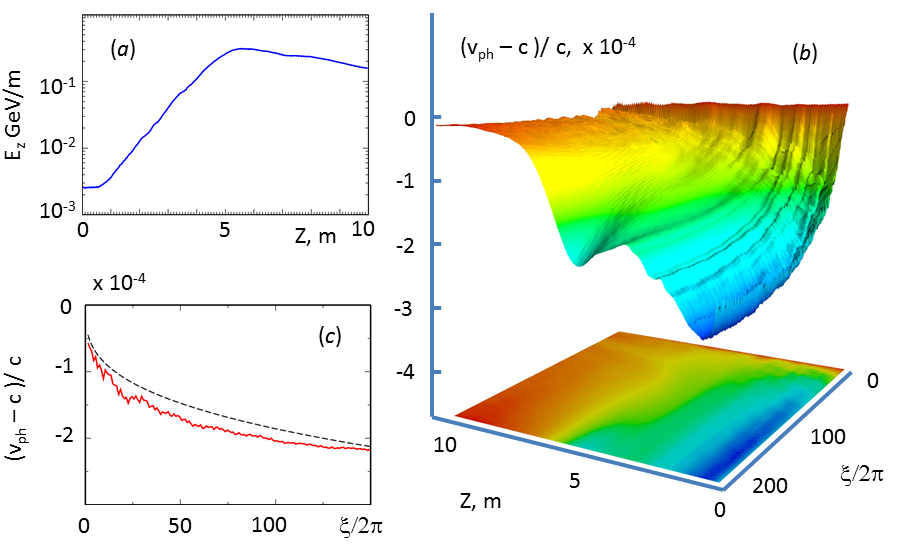}

\caption{\label{fig:(color-online)-3D}(color online) A 3D hybrid PIC simulation
of SPS 
bunch self-modulation in plasma with constant density $n_{e}=7.76\cdot10^{14}$
cm$^{-3}$. Frame (a) shows the evolution of the maximum accelerating
field. The linear instability persists for some 5 meters before the
nonlinear saturation. Frame (b) shows the phase velocity of the wake
as a surface dependent on the propagation distance $z$ and the bunch
coordinate $\xi$. The phase velocity experiences a deep drop towards
the tail of the bunch at the linear stage of the instability. When
the bunch is completely modulated and the instability saturates at
$z>5$m, the wake phase velocity surges up to that of the driving
bunch. Frame (c) gives a snapshot of the wake phase velocity after
2.5 meters of propagation distance (the solid red line). The broken
line gives the analytical expression \eqref{eq:Vph}. }
\end{figure}

The simulation results are presented in Fig. \ref{fig:(color-online)-3D}.
From the frame (a) we see that the linear instability stage lasts
during the first 5 meters of propagation. Then, the bunch is completely
modulated and the nonlinear saturation is reached. The wake phase
velocity is shown in the frame (b). A significant slowdown of the
wake is observed during the instability and along the bunch. The lowest
phase velocity is The lowest relative phase velocity is ($\upsilon_{\mathrm{ph}}-c)/c\approx-3.5\cdot10^{-4}$,
corresponding to a wake $\gamma_{ph}\approx38$. This is an order
of magnitude lower than the $\gamma-$factor of the driving bunch.
Frame (c) compares the simulation result (the solid red curve) with
the analytic expression \eqref{eq:Vph}, in which we substituted the
SPS bunch parameters. This snapshot of the wake phase velocity has
been taken at $z=2.5\,$m, in the middle of the linear instability
stage. A reasonable agreement between the simulation and the analytical
theory is observed.

The wake slowdown has a dramatic impact on the electron trapping and
acceleration. First, it allows for trapping of low energy electrons
whose velocities are comparable with the wake velocity. However, the
energy gain in the slow wake is very limited due to fast dephasing.
The energy gain is given by $\Delta W\approx4\gamma_{ph}^{2}mc^{2}(E_{max}/E_{WB})$
\cite{esarey review}. At the linear stage we have $E_{max}\ll E_{WB}$
and the energy gain is low for small $\gamma_{ph}$. The dephasing,
however, has a much worse effect : if the electrons have been injected
into the early instability phase of the slow wake, they can be lost
when they overtake the wave and enter its defocusing phase. The dephasing
distance is $k_{p}L_{d}\approx2\pi\gamma_{ph}^{2}$ and for the slow
wake field it can be shorter than the distance needed for the instability
to develop. For this reason, the electrons must be injected late in
the development of the instability, when the phase velocity begins
to grow. In our simulation, the optimum point for injection is located
around $z=5\,$m. The wake phase velocity is still low here, but starts
growing rapidly as the bunch reaches complete modulation.

A possibility to inject electrons into the wake is side injection
\cite{side injection}. In this case, a bunch of electrons is propagating
at a small angle with respect to the driver. The advantage of side
injection over on-axis injection is that electrons are gradually {}``sucked-in''
at the right phase by the wake transverse field. This leads to high
quality quasi-monoenergetic acceleration of electrons.

\begin{figure}
\includegraphics[width=0.5\columnwidth]{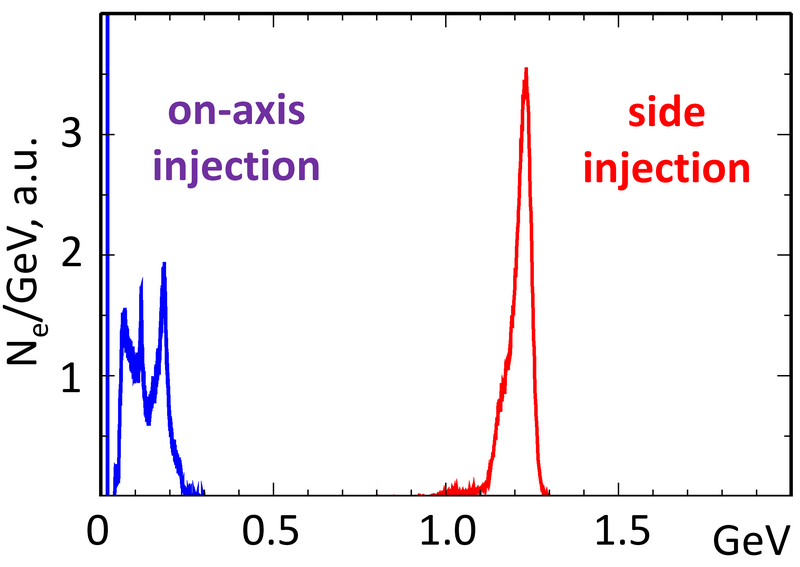}

\caption{\label{fig:Electron-energy-spectra}(color online) Electron energy
spectra at $z=10\,$m. There were two groups of electrons. One was
injected on-axis at the plasma entrance; another was side-injected
at the angle of 0.005 radian. In the latter case, the electron bunch
trajectory should cross the driver axis at $z=6\,$m, $\xi=20\,$cm
behind the bunch head. The on-axis injection lead to a broad spectrum
and low energy gain. The side injection resulted in a quasi-monoenergetic
beam at 1.2 GeV energy and about 1\% energy spread. Initial electron
energy was 20 MeV for on-axis injection and 10 MeV for side injection.}
\end{figure}

In our simulation, we have injected two bunches of test electrons.
The first bunch of 20 MeV electrons was injected on-axis at the plasma
entrance. This electron energy roughly corresponded to the minimum
wake phase velocity at the tail of the driver. We found that during
the linearly growing instability stage, these electrons underwent
more than one oscillation in the ponderomotive bucket. Finally, after
10 meters of propagation, the maximum energy gain was about 200 MeV
with a rather broad energy spectrum as seen in Fig. \ref{fig:Electron-energy-spectra}.

\begin{figure}
\includegraphics[width=0.9\columnwidth]{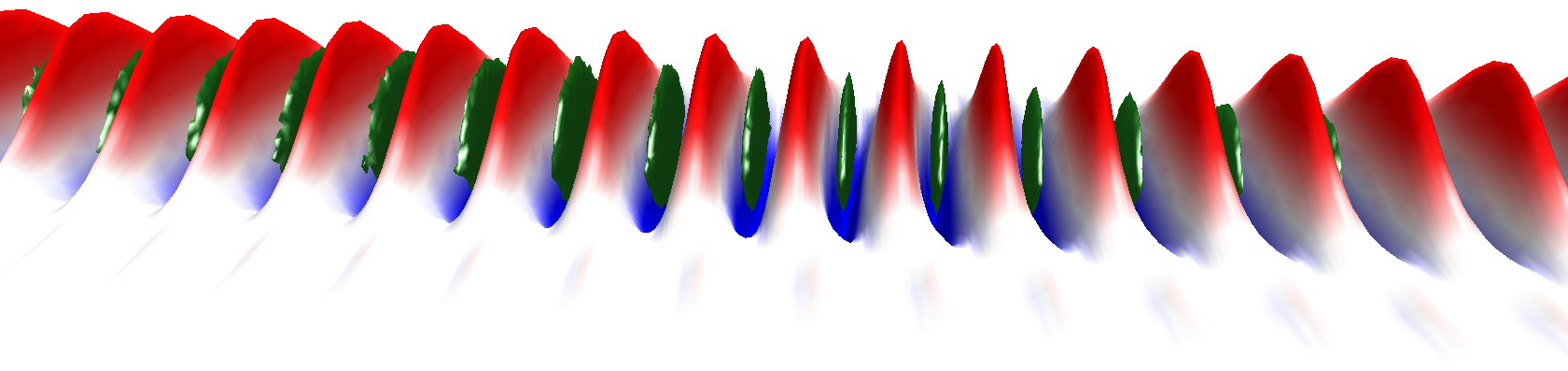}

\caption{\label{fig:Test-electron-beam}(color online) Test electron bunch
(dark green discs) and the accelerating field (the red-blue wave)
at $z=5\,$m. The electrons are captured and split into micro bunches
located exactly in the accelerating and focusing phases of the wake.}
\end{figure}

The second electron beam with 10 MeV energy was side-injected at an
angle of 0.005 radian. The electron bunch trajectory was designed
to cross the driver axis at $z=6\,$m, $\xi=20\,$cm behind the bunch
head. Due to the small injection angle, however, the electrons were
sucked in into the wake much earlier, at the position $z\approx5$m.
The wake transverse fields have put most of the bunch electrons into
the focusing and accelerating phase. The electron beam and field configuration
just after the electrons entered the wake is shown in Fig. \ref{fig:Test-electron-beam}.
The electron beam is split into micro-bunches located exactly in the
accelerating and focusing phases of the wake. Due to this configuration,
the side injected beam resulted in a maximum energy gain of 1.2 GeV
and a rather narrow energy spectrum.

The low energy spread and efficient acceleration of the side injected
electrons are also due to the fast rise in the wake phase velocity
just after the injection position, as seen in Fig. \ref{fig:(color-online)-3D}(b).
The electrons gain energy while staying in the accelerating phase
of the wake.

\begin{figure}
\includegraphics[width=0.9\columnwidth]{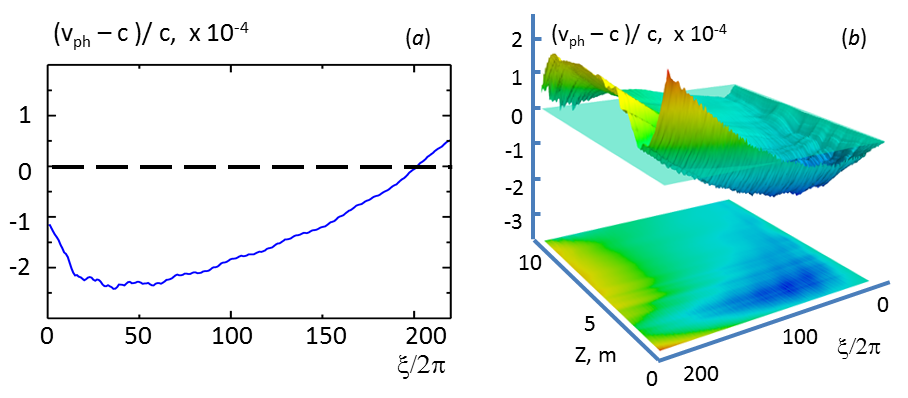}

\caption{\label{fig:Control of Vph} (color online) Phase velocity control
by positive plasma density gradient. a) Phase velocity at \textit{z}=2.5
m. b) Full dynamics picture; the transparent plane marks the speed
of light. }
\end{figure}

We have seen above that the growing mode \eqref{eq:unstable modes}
has the low phase velocity \eqref{eq:Vph}. There is a way, however,
to manage the phase velocity of the wake by employing a gentle longitudinal
plasma density gradient as it was discussed in \cite{pukhov kostyukov}.
To elucidate the effect, we have performed an additional simulation
with the same beam parameters, but introducing a positive plasma density
gradient: $n_{e}(z)=n_{e0}(1+z/d)$ with the characteristic length
$d=200$m. The phase velocity obtained in this simulation is shown
in Fig. \ref{fig:Control of Vph}.

The phase velocity at the head of the beam takes a dive as defined
by the growing mode dispersion. However, the positive plasma density
gradient compensates for the mode dispersion and at the tail of the
beam the wake phase velocity becomes equal to the speed of light and
even superluminous.

In summary, we have shown that the self-modulational instability of
a charged beam in plasma corresponds to a growing mode with a slow
phase velocity. The wake velocity is much lower than that of the driver.
The wake slowdown is due to the real part of the frequency of the
unstable mode. Although this effect limits electron energy gain at
the stage of the linear instability growth, the low phase velocity
can be harnessed to inject low energy electrons into the wake of a
highly relativistic driver. We also have shown that the side injection
of electrons at a small angle with respect to the driver axis may
drastically improve the quality of acceleration. The transverse field
of the wake sucks in the injected electrons and automatically puts
them into the right acceleration phase. Finally, we show that the
wake phase velocity can be controlled by a longitudinal plasma density
gradient.

This work was supported by DFG. V.K., C.~S., and G.~S. acknowledge
funding support from the US Department of Energy under the grant DE-FG02-04ER41321.


\begin{thebibliography}{References}
\bibitem{tajima dawson} T. Tajima and J. M. Dawson, Laser Electron
Accelerator, Phys. Rev. Lett. \textbf{43}, 267\textendash{}270 (1979).

\bibitem{joshi malka} Chan Joshi and Victor Malka \textquotedbl{}\emph{Focus
on Laser- and Beam-Driven Plasma Accelerators}\textquotedbl{}. New
Journal of Physics. (2010); C. Joshi et al., Phys. Plasmas \textbf{9},
1845\textendash{}1855 (2002).

\bibitem{joshi scientific american}{]} C. Joshi, \textquotedbl{}Plasma
Accelerators,\textquotedbl{} Scientific American (February 2006),
\textbf{294}, 40-47

\bibitem{TR} K. L. Bane, P. Chen, and P. B. Wilson, IEEE Trans. Nucl.
Sci. \textbf{32}, 3524 (1985).

\bibitem{pwa}{]} A. Caldwell, K. Lotov, A. Pukhov, and F. Simon,
Nat Phys, \textbf{5}, 363 (2009);

\bibitem{lotov}K.V.Lotov Phys. Rev. ST-AB, \textbf{13 }041301 (2010).

\bibitem{esarey sm} E. Esarey, J. Krall, and P. Sprangle, Phys. Rev.
Lett., \textbf{72}, 2887 (1994).

\bibitem{bubble} A. Pukhov and J. Meyer-ter Vehn, Applied Physics
B: Lasers and Optics, \textbf{74}, 355 (2002).

\bibitem{kumar sm} N. Kumar, A. Pukhov, K. Lotov. Phys. Rev. Lett.
\textbf{104}, 255003 (2010).

\bibitem{esarey review} E. Esarey, et al. Rev. Mod. Phys. \textbf{81},
1229 (2009).

\bibitem{andreev vph} N. E. Andreev et al., IEEE TPS \textbf{24},
363 (1996).

\bibitem{bers} A. Bers, {}``Basic plasma physics 1,\textquotedblright{}
(North-Holland Publishing Company, 1983) Chap. 3.2.

\bibitem{Lotov2011} K.V.Lotov, Phys. Plasmas \{\textbackslash{}bf
18\} 024501 (2011). 

\bibitem{tobi2010} T. Tueckmantel et al, IEEE TPS \textbf{38}, 2383
(2010).

\bibitem{side injection} S. Yu. Kalmykov et al., Phys. Plasmas \textbf{13},
113102 (2006).

\bibitem{pukhov kostyukov}A. Pukhov, I. Kostyukov, Phys. Rev. E \textbf{77},
025401 (2008).

\bibitem{whittum_prl91} D. Whittum \textit{et. al.,} Phys. Rev. Lett.\textbf{
67,} 991 (1991).\end{thebibliography}
\end{document}